\documentclass[]{article}
\usepackage{lmodern}
\usepackage{amssymb,amsmath}
\usepackage{ifxetex,ifluatex}
\usepackage{fixltx2e} % provides \textsubscript
\ifnum 0\ifxetex 1\fi\ifluatex 1\fi=0 % if pdftex
  \usepackage[T1]{fontenc}
  \usepackage[utf8]{inputenc}
\else % if luatex or xelatex
  \ifxetex
    \usepackage{mathspec}
    \usepackage{xltxtra,xunicode}
  \else
    \usepackage{fontspec}
  \fi
  \defaultfontfeatures{Mapping=tex-text,Scale=MatchLowercase}
  
\fi
\IfFileExists{upquote.sty}{\usepackage{upquote}}{}
\IfFileExists{microtype.sty}{%
\usepackage{microtype}
\UseMicrotypeSet[protrusion]{basicmath} % disable protrusion for tt fonts
}{}
\usepackage[margin=1in]{geometry}
\usepackage{longtable,booktabs}
\usepackage{graphicx}
\makeatletter
\def\maxwidth{\ifdim\Gin@nat@width>\linewidth\linewidth\else\Gin@nat@width\fi}
\def\maxheight{\ifdim\Gin@nat@height>\textheight\textheight\else\Gin@nat@height\fi}
\makeatother
\setkeys{Gin}{width=\maxwidth,height=\maxheight,keepaspectratio}
\ifxetex
  \usepackage[setpagesize=false, % page size defined by xetex
              unicode=false, % unicode breaks when used with xetex
              xetex]{hyperref}
\else
  \usepackage[unicode=true]{hyperref}
\fi
\hypersetup{breaklinks=true,
            bookmarks=true,
            pdfauthor={Khushnood Abbas\^{}\{1\},201414060110@std.uestc.edu.cn; Shang Mingsheng\^{}\{2\} ,msshang@cigit.ac.cn; Luo Xin\^{}\{3\} ,luoxin21@cqu.edu.cn; 1.Web science center University of Electronic science and Technology Of China (UESTC); 2.Chongqing Institute of Green and Intelligent Technology, Chinese Academy of Sciences(CIGIT); 3.Chongqing University Chongqing China},
            pdftitle={Discovering items with potential popularity on social media},
            colorlinks=true,
            citecolor=blue,
            urlcolor=blue,
            linkcolor=magenta,
            pdfborder={0 0 0}}
\let\rmarkdownfootnote\footnote%
\def\footnote{\protect\rmarkdownfootnote}

\usepackage{titling}

  \title{Discovering items with potential popularity on social media}
  \pretitle{\vspace{\droptitle}\centering\huge}
  \posttitle{\par}
  \author{Khushnood
\(Abbas^{1}\),\href{mailto:201414060110@std.uestc.edu.cn}{\nolinkurl{201414060110@std.uestc.edu.cn}} \\ Shang \(Mingsheng^{2}\)
,\href{mailto:msshang@cigit.ac.cn}{\nolinkurl{msshang@cigit.ac.cn}} \\ Luo \(Xin^{3}\)
,\href{mailto:luoxin21@cqu.edu.cn}{\nolinkurl{luoxin21@cqu.edu.cn}} \\ 1.Web science center University of Electronic science and Technology Of
China (UESTC) \\ 2.Chongqing Institute of Green and Intelligent Technology, Chinese
Academy of Sciences(CIGIT) \\ 3.Chongqing University Chongqing China}
  \preauthor{\centering\large\emph}
  \postauthor{\par}
  \date{}
  \predate{}\postdate{}

\begin{document}

\maketitle

\section{Abstract}\label{abstract}

Predicting the future popularity of online content is highly important
in many applications. Preferential attachment phenomena is encountered
in scale free networks.Under it's influece popular items get more
popular thereby resulting in long tailed distribution problem.
Consequently, new items which can be popular (potential ones), are
suppressed by the already popular items. This paper proposes a novel
model which is able to identify potential items. It identifies the
potentially popular items by considering the number of links or ratings
it has recieved in recent past along with it's popularity decay. For
obtaining an effecient model we consider only temporal features of the
content, avoiding the cost of extracting other features. We have found
that people follow recent behaviours of their peers. In presence of fit
or quality items already popular items lose it's popularity. Prediction
accuracy is measured on three industrial datasets namely Movielens,
Netflix and Facebook wall post. Experimental results show that compare
to state-of-the-art model our model have better prediction accuracy.

\section{Introduction}\label{introduction}

E-commerce and social-media offer their users facilities to buy, review,
sell and share online items, such as Amazon, e-Bay, Facebook and Tencent
\href{https://en.wikipedia.org/wiki/Tencent_QQ}{QQ}. The adoption of
social media is increasing by day and night so as e-commerce. Because of
its widely adoption, it is gaining attention of the organizations, to
know people's demand and trend.Researchers have also found correlation
between social media and e-commerce industry. Social media is showing a
great opportunity by carrying huge amount of user generated data for
exploiting and finding the outcome of interest (Asur, Huberman, and
others 2010). In e-commerce site, organizations are directly interested
what are the response from users, for their products. They also want to
know which kind of item will be in demand so that they can make profit.
Since the number of products are expanding constantly so online
merchants changes their strategies from traditional marketing
advertisement (like TV, news paper etc) to viral marketing,
i.e.~customers are suggested to share the product information with their
friends on social media such as Facebook and twitter (Leskovec, Adamic,
and Huberman 2007). From infotainment to trade, everything is done on
Internet and online contents have become valuable internet asset (Tatar
et al. 2014) that is useful to producers and consumers both, hence to
know the future popularity of online content has become an important
area of attention and interest.

Popularity prediction is a complex task depending on different factors
like quality and individual's interest.Content popularity may fluctuate
with time (Eisler, Bartos, and Kert{\a'e}sz 2008), increase over time or
be limited within communities. It is difficult to capture the
relationship between real world event and web content, in the prediction
model for example during Indian election many joke goes viral and they
might not explicity inlude the political symbol.Some contents get
extreme popularity because of its prior popularity, also known as
\emph{cascading effect} (Cheng et al. 2014), and it becomes hard to
predict which content will stop this cascading effect. In presence of
cascading effect other ``potential items'' (exhibit a sudden increase of
popularity at a certain duration of time, i.e.~can be popular, but not
now) are suppressed.Most recently researchers have found that the
popularity of online contents like news , blog posts , videos, mobile
app download (Gleeson et al. 2014) in online discussion forums and
product reviews exhibits temporal dynamics. It turns out that user
interests toward web items, vary with time; an extensive study of how
the popularity of online media's temporal patterns grow and fade, over
time are presented in (Yang and Leskovec 2011), (Leskovec, Backstrom,
and Kleinberg 2009). Besides the works on temporal scale, other
researchers found that the contents popularity is influenced
significantly by consumers social relationship (ZENG et al. 2013),
(Szabo and Huberman 2015), the social network of consumer can enhance
the prediction performance; and social dynamics of consumer, influence
the social media content's popularity even more significantly. Because
the models of both temporal dynamics and social dynamics are always
complex and parameterized, it is hard to apply those models to real
online systems.

The rest of the paper is organized as follows. In Section 2 we formally
defined problem. In section 3 we introduced the baseline method.
Insection 4 we proposed new method to solve the problem. In section 5 we
discuss the mthods and materials for experiment and we also discuss
insights from the results.In Section 6 we have concluded the paper with
possible future works.

\section{Problem definition}\label{problem-definition}

In our model we have considered the bipartite network which consists of
a set of users (\(U\)) and a set of objects (\(O\)). The popularity of a
node or item or object is the total link recieved by the object (\(o\)).
A bipartite network can be represented by an adjacency matrix \(A\), if
user \(u\), (\(u \in U\)) have consumed object \(o\), (\(o \in O\)). To
consider the temporal effect on objects' final popularity we take
snapshots of the network at different time point. Let \(A(t)\) denote
the adjecency matrix of the snap-shotted network at time \(t\), then
matrix (\(A(t)\)) containes edges between user (\(u\)) and object
(\(o\)) before time (\(t\)) only. The user and object degree can be
computed by \(k_u (t) = \sum\limits_o {A_{uo} (t)}\) and
\(k_o (t) = \sum\limits_u {A_{uo} (t)}\), respectively.\(T_F\) is future
time window, so popularity or increment in degree of an object \(o\) is
given by-

\[
\begin{aligned}
\Delta k_o (t,T_F ) = k_o (t,T_F ) {-} k_o (t)
\end{aligned}
\]

The popularity prediction problem can be defined as follows. Given a
data-set S which includes information about ratings by users that have
rated or consumed the item/object \(o\) with time-stamp (e.g.~user id,
item id, time), after arranging the data-set in ascending order by time,
we divided the data-sets into training time t and future time window
\(T_F\). This is obvious that interest towards object, varies with type
or category such as online video, object is different than items on
Amazon. Considering these facts we have chosen the training time \(t\)
that helps us to predict with better accuracy in future time window. A
predictor exploits the information before time (\(t\)) and make
prediction for future time window (\(T_F\)). \(\Delta k_o (t,T_F )\) is
the real score of object (\(o\)). A predictor's performance is measured
by calculating it's accuracy by comparing its ranking from predicted
ranking and real ranking.

\section{Baseline method for
comparison}\label{baseline-method-for-comparison}

We have considered state-of-the art for predicting new entries as a
baseline method as for as our knowledge goes.

\subsection{Popularity-based
predictor}\label{popularity-based-predictor}

(ZENG et al. 2013) has proposed Popularity based predictors(PBP). It is
based on a well known \emph{preferential attachment} theory , which
states that popularity increases cumulatively; the rate of new link
(Either item recieves rating in case of Movielens, or a friend like or
comments in case of Facebook wall post activity) formation for any node
is proportional to the observed number of links which node has recieved
in past. If an item is popular at time \({\rm t}\), then it will
probably become popular due to the condition that current degree of an
item \({\rm k}_o {\rm (t} {\rm )}\) is a good predictor of its future
popularity. Further ((Gleeson et al. 2014), (ZENG et al. 2013)) have
found that current degree is a good predictor of items' future
popularity.(ZENG et al. 2013) proposes to calculate the prediction score
of an item at time \(t\) can be given as follows- -

\begin{equation}
{\rm s}_o  {\rm (t} {\rm ,T}_{\rm p} {\rm ) = k}_o  {\rm (t} ) - \lambda {\rm k}_o  {\rm (t} {\rm , T}_{\rm P} {\rm )}
\end{equation}

Where \(\Delta {\rm k}_o {\rm (t} {\rm , T_P)}\) is the rating/links
recieved in past time window \(T_P\) from \(t\).
\(\lambda \in {\rm [0, 1] }\) , note that \(\lambda = 0{\rm }\) gives
the total popularity and for \(\lambda = 1{\rm }\) it gives recent
popularity. Throught the script by popularity we mean number of ratings
or links recieved by item or node.

\section{Proposed method: Considering aging factor with recent
popularity.}\label{proposed-method-considering-aging-factor-with-recent-popularity.}

It is obvious that popularity of any item on social media doesn't last
forever.In addition decay rate may vary from item to item e.g life cycle
of popularity of a movie will be different from news or items on Digg or
Facebook.Every type of item have its own decay rate like (Parolo et al.
2015) have found research article citation rate decays after some time.
Mathew effect or preferential attachement is a well known phenomena seen
almost in large scale networks that shows power law distribution (et.al
1999). These theories explains that rich will get richer and poor get
poorer.

As researchers have found that degree distribution of every network ends
up to long tailed.This is also true in case of e-commerce user-item
bipartite networks.People chase the popularity of items to optimize
their time and energy. There are good or fit items ({Matus Medo Manuel
S. Mariani} and Zhang 2015), (Bianconi and Barab{Ã}{\textexclamdown}si
2001) to be consumed but under the influence of preferential attachment
those are ignored. Under it's influence finding the fit or potential
item is one of the important tasks among the researchers.Researchers
have also found that network changes structure due to aging factor over
time {[}(H. Zhu, Wang, and Zhu 2003){]}. Considering these factors we
propose that the recent gain in popularity as well as decay in
popularity together are a good predictor for its future
popularity.Considering the competing behavior in networks (Bianconi and
Barab{Ã}{\textexclamdown}si 2001),if some items are loosing their
populairty then other items should be gaining attentions of the
consumer. Therefore decaying factors with recent popularity will help us
in detecting ``potantial items''. Recent popularity is one of the
important factors in discovering the final popularity of objects are
discussed in (J.-P. Onnela and Reed-Tsochas 2010). We also know that
considering all features that affect popularity of content is really a
difficult task. If \({\rm s}_o {\rm (t} {\rm ,T}_{\rm p}{\rm )}\) is
prediction score at time \(t\) given past time window \(T_P\). We can
say-

\begin{equation}
s_o (t ,T_p ) \prec \sum\limits_u {(k_o (t) - \lambda k_o (t - T_P )) }
\end{equation}

The above equation states that score of object is proportional to recent
gain in popularity. \(\lambda\) is tunable prameter between recentness
and total popularity.it can take values in {[}0,1{]} interval. As the
researchers also found aging phenomina in item or node so we can
formulate it as follows-

\begin{equation}
s_o (t ,T_p ) \prec \sum\limits_u { e^{\gamma (T_{uo}  - t)} }
\end{equation}

Where \(T_{uo}\) denotes the time at which user \(u\) consumed the
object \(o\) and \(\gamma\) is free parameter.Since recent popularity
will be good predictor if decay rate is constant. So now we can write as
follows-

\begin{equation}
s_o (t ,T_p ) \prec \sum\limits_u {(k_o (t) - \lambda k_o (t - T_P ))*e^{\gamma (T_{uo}  - t)} }
\end{equation}

again we can write-

\begin{equation}
s_o (t ,T_p ) = l\sum\limits_u {(k_o (t) - \lambda k_o (t - T_P ))*e^{\gamma (T_{uo}  - t)}} 
\end{equation}

where \(l\) is normalization constant and can be estimated using
following equation \(\sum\limits_o {s_o (t ,T_p )} = 1\).

\section{Experimental Results and
discussion}\label{experimental-results-and-discussion}

For testing our proposed predictor's efficiecy we have considered
Popularity Based Predictor (PBP) as a base predictor by (ZENG et al.
2013). We took average of 10 results.

\subsection{Evalutation metrics}\label{evalutation-metrics}

Three evaluation metrics are adopted to measure the accuracy of the
proposed model including
\emph{precision}\((P_n)\),\emph{novelty}\((Q_n)\) and \emph{Area Under
Recieving Operating Characteristic}(\(AUC\)).

\begin{itemize}
\itemsep1pt\parskip0pt\parsep0pt
\item
  \emph{Precision} is defined as the fraction of objects that are
  predicted also lie in the top \(N\) object of true ranking (Herlocker
  et al. 2004).
\end{itemize}

\[
\begin{aligned}
 {p_n  = \frac{{D_n }}{n}}
\end{aligned}
\] Where \(D_n\) is the number of common objects between predicted and
real ranking. \(n\) is the size of list to be ranked.It's value ranges
in {[}0,1{]}, higher value of \((P_n)\) is better.

\begin{itemize}
\itemsep1pt\parskip0pt\parsep0pt
\item
  \emph{Novelty(\(Q_n\))} is a metric to measure the ability of a
  predictor to rank the items in top \(n\) position that was not in top
  \(n\) position in previous time window.We call these new entries as
  ``potential items'' throughout the script. If we denote the predicted
  object as (\(P_po\)) and potential true object as \(P_ro\), then the
  novelty of a model is given by-
\end{itemize}

\[
\begin{aligned}
Q_n  = P_{po} /P_{ro}
\end{aligned}
\]

\begin{itemize}
\itemsep1pt\parskip0pt\parsep0pt
\item
  \emph{AUC} measures the relative position of the predicted item and
  true ranked items. Suppose predicted item list is (\(L_pn\)) and real
  item list is (\(L_rn\)). if \(s_op \in L_{pn}\) and
  \(s_rp \in L_{rn}\) is score of object in predicted then \emph{AUC} is
  given by-
\end{itemize}

\[
\begin{aligned}
AUC = \frac{{\sum\limits_{op \in L_{pn} } {\sum\limits_{rp \in L_{rn} } {I(s_{pn} ,s_{rn} )} } }}{{\left| {L_{pn} } \right|\left| {L_{rn} } \right|}}
\end{aligned}
\]

where, \[
\begin{aligned}
I(s_{pn} ,s_{rn} ) = \left\{ 
   {\begin{array}{*{20}c}
   {0 \Leftarrow s_{pn}  < s_{rn} }  \\
   {0.5 \Leftarrow s_{pn}  < s_{rn} }  \\
   {1 \Leftarrow s_{pn}  > s_{rn} }  \\
\end{array}} \right.
\end{aligned}
\]

\subsection{Data used in this article}\label{data-used-in-this-article}

To test the predictors accuracy we have used different data sets. Like
MovieLens, Netflix, Facebook wall post datasets etc. MovieLens and
netflix data sets contain movie ratings and Facebook data set contains
users' wall post relationships. MovieLens is provided by
\href{www.grouplens.org}{GroupLens} project at University of Minnesota.
The data description can be found on the website.While data preparation
for our model we have selected small subset from each by randomly
choosing users who have rated atleast \(20\) movies. The original rating
was in the form of numarical \(1-5\), we have considered the link
between the user and object which object have recieved higher than two
ratings.For all the three datasets Facebook, Movielens and Netflix the
time is considered in days. The data description is as follows-

\begin{itemize}
\item
  \textbf{Netflix} data contains \(4960\) users,\(16599\) movies and
  \(1249058\) links, data was collected during(1st Jan \(2000\)
  --\(31st\) Dec 2005).
\item
  \textbf{MovieLens} dataset contains \(7533\) movies, \(864581\) links
  and \(5000\) users and data was collected during(\(1st\) Jan \(2002\)
  --\(1st\) Jan 2005).
\item
  \textbf{Facebook} data contains \(40981\) set of users and their
  \(38143\) wall post activity and \(855542\) links, during period of
  (14 Sep \(2004\)--\(22nd\) Jan \(2009\)). If user has posted on a wall
  there will be a link between the user and the wall, self influenced is
  removed by removing the link between user and its own wall post.
\end{itemize}

\subsection{Accuracy results on different
datasets}\label{accuracy-results-on-different-datasets}

To evaluate the performance of our predictors we have selected 10 random
\(t\) for each data sets. Selection of t is considered in such a way
that predictor have enough history information. Since predictors are
based on objects' history, we have selected only those object that have
recieved atleast one link before time \(t\).

\includegraphics{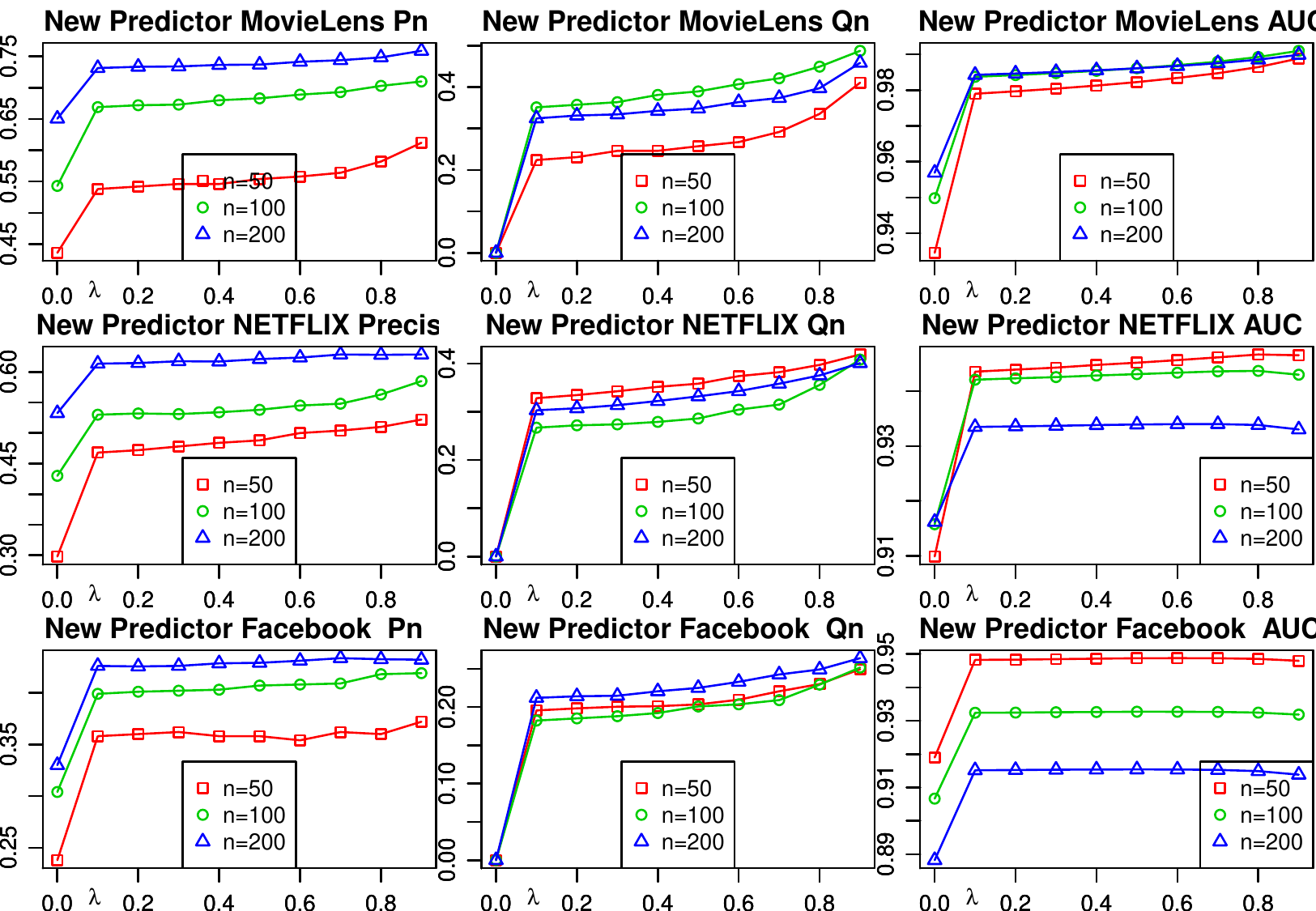}

\begin{quote}
Figure 1: The performance of the proposed method for different values of
\(n\) (Top n items in a list). For Three data sets Movielens, Netfix and
Facebook wall post.
\end{quote}

In the above {[}Figure 1{]} we have shown the proposed predictors
performance under different values of \(\lambda\), i.e considering the
recent popularity with variying history lenght because \(\lambda=0\)
means objects total popularity and \(\lambda=1\) means objects recent
popularity. We have consider the decay rate \(\gamma\) as fix, because
we have found \(\gamma=0.1\) gives better result.From the figure
{[}Figure 1{]} we can see \(P_n\) is much affected by \(n\) than \(Q_n\)
(novelty) and \(AUC\). All the three metrics improves with \(\lambda\)
in other words recent populairty is a good predictor than the object's
total popularity. Higher \(P_n\) values shows that proposed predictor
have better ability to predict popularity of objects than the base
method while our predictor have ability to predict better the popularity
of novel items(\(Q_n\)),i.e the items that were not popular in the past
time. These items are ``potential items'', these items will help in
abating the centrality of item's degree distribution. Generally these
items are suppressed by items that have already gained populairty. Our
proposed predictor have also shown improvment over the base method.

\subsection{Effect of aging Vs recent
popularity}\label{effect-of-aging-vs-recent-popularity}

In decay rate (\(\gamma\)) and recent popularity analysis we have found
decay rate is very low for all the three datasets. Considering decay
with recent populartiy improves accuracy. We have also found considering
decay helps more in digging new entries. Although aging factor improves
the accuracy but still recent behavior dominates. We have also found
that in the presence of quality item people lose interest in old items
that is why \(Q_n\) improves when considering decay factor with recent
poularity.Even if the item is not globally popular people like the items
that were liked by the peers recently. Empirical results are as follows-

\includegraphics{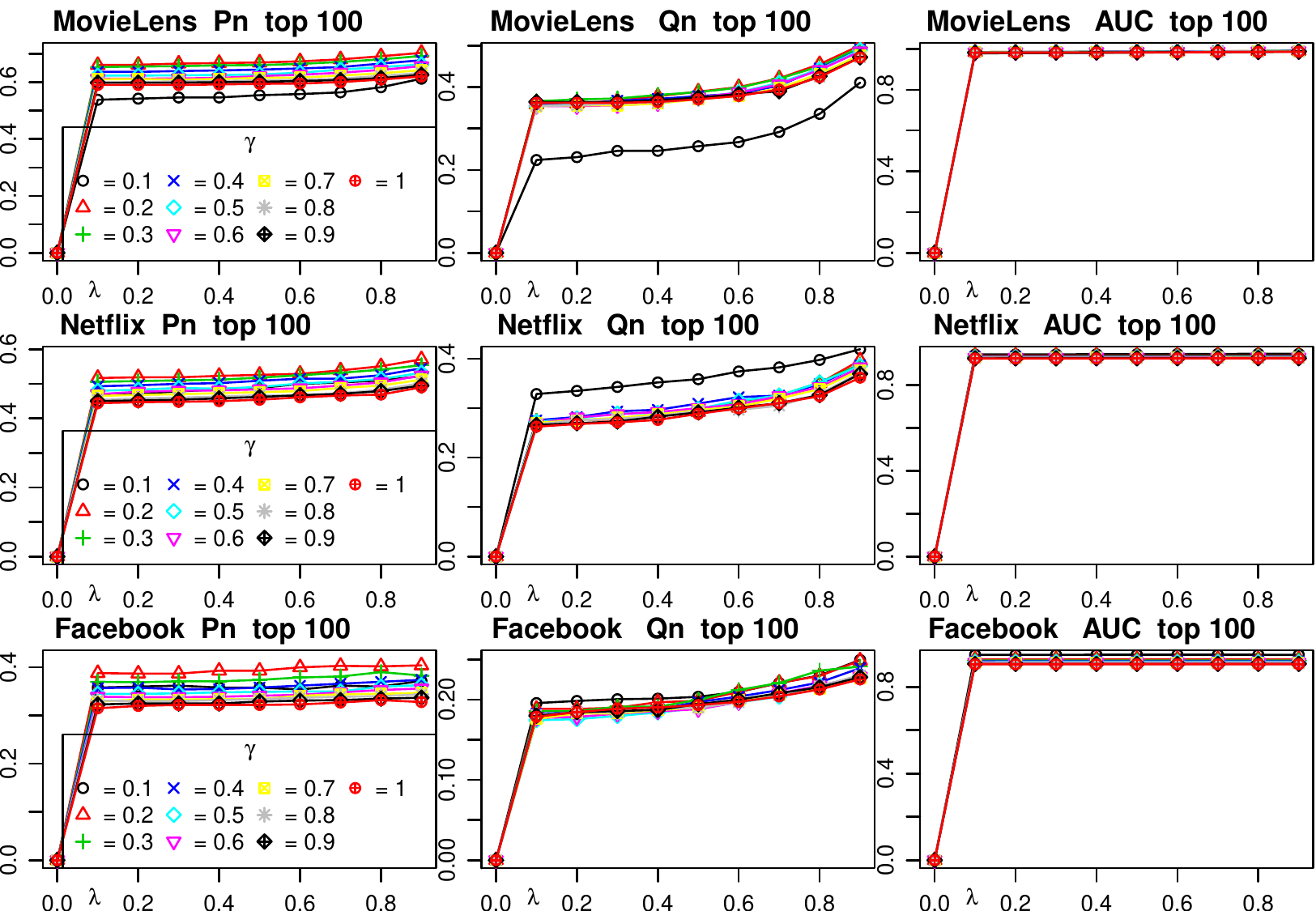}

\begin{quote}
Figure 2: The performance of our proposed predictor for different values
of \(\lambda\) and \(\gamma\) (decay rate).We have considered \(T_F\)
and \(T_P\) both as \(30\) days.
\end{quote}

In search of our predictor's behavior we have considered different
values of \(\gamma\) and plotted {[}Figure 2{]} the accuracy against
\(\lambda\). Although \(\gamma\) is free parameter but to know its
effect we considered 10 values in {[}0,1{]}. We have tried recentness of
edge collection by the node with decay rate to see the effect together.
We have found popular items last longer on Movielens since precision is
better for \(\gamma=0.2\). We have found people adopts new items from
recent behaviours of their peers. In other words for trying new things
they rely on their peer's behaviour.Since \(Q_n\) not much affected by
decay rate as compare recent behaviours. In cas of Netflix we found the
same nature as Movielens popular items last for long time. In case of
Facebook we have found that people rely more on the age of the post than
the recent behaviours of their friends. While more recent post shared by
their friends are more entertained.On Facebook not only share or comment
on their friend's post but also they can create their own post and other
can share or comment.That is why potantial items on Facebook,a user not
only depends on the age and recent activity of the peers but also node
cetrality. Who have shared the post also matters.That is why predicting
new popular items on Facebook needs more feature consideration such as
centrality of the node,time etc.

The {[}Table 1{]} gives detailed comparison of the two predictors. We
have considered future time window as well as past time window both as
\(30\) days. The second column is accuracy of our proposed predictor
while PBP column is the base predictor.The three numbers in both the
columns are basically for top 50,100 and 200 items respectively from
left to right. The ``Type'' column describe the accuracy type, every
accuracy is compared for three cases : n=50,100 and 200 items.It is easy
to see that proposed predictor has shown improvement.

\subsection{Comparison of two
predictors}\label{comparison-of-two-predictors}

For comparing our proposed predictor with base predictor we have
considered past time window (\(T_P\)) and future time window (\(T_F\))
as \(30\) days. For comparison we have selected the top n ranked items
from predicted list and compare them against the real items for both the
predictors. Emprical results are as follows-

\begin{longtable}[c]{@{}llll@{}}
\caption{Perfomrance table for both the predictor considering Tp and Tf
as 30 days. The volues are for top 50,100 and 200 items
respectively.}\tabularnewline
\toprule
DataSet & Proposed 50/100/200 & PBP 50/100/200 & Type\tabularnewline
\midrule
\endfirsthead
\toprule
DataSet & Proposed 50/100/200 & PBP 50/100/200 & Type\tabularnewline
\midrule
\endhead
MovieLens & 0.612, 0.71, 0.759 & 0.498, 0.646, 0.721 & Pn\tabularnewline
Netflix & 0.522, 0.585, 0.628 & 0.462, 0.534, 0.609 & Pn\tabularnewline
Facebook & 0.372, 0.419, 0.434 & 0.358, 0.398, 0.426 & Pn\tabularnewline
MovieLens & 0.41, 0.487, 0.458 & 0.126, 0.253, 0.243 & Qn\tabularnewline
Netflix & 0.419, 0.408, 0.401 & 0.302, 0.263, 0.267 & Qn\tabularnewline
Facebook & 0.249, 0.252, 0.264 & 0.19, 0.177, 0.207 & Qn\tabularnewline
MovieLens & 0.989, 0.991, 0.99 & 0.976, 0.982, 0.982 &
AUC\tabularnewline
Netflix & 0.947, 0.944, 0.934 & 0.943, 0.942, 0.936 & AUC\tabularnewline
Facebook & 0.949, 0.933, 0.915 & 0.947, 0.933, 0.916 &
AUC\tabularnewline
\bottomrule
\end{longtable}

\subsubsection{Accuracy comparison}\label{accuracy-comparison}

We have compared our results with the base method considering top
50,100,200 list. For comparing we have considered training windown
(\(T_P\)) as 30 days and we have tested the predictor for the same
future time lenght \(T_F = 30\) days.

\begin{center}\includegraphics{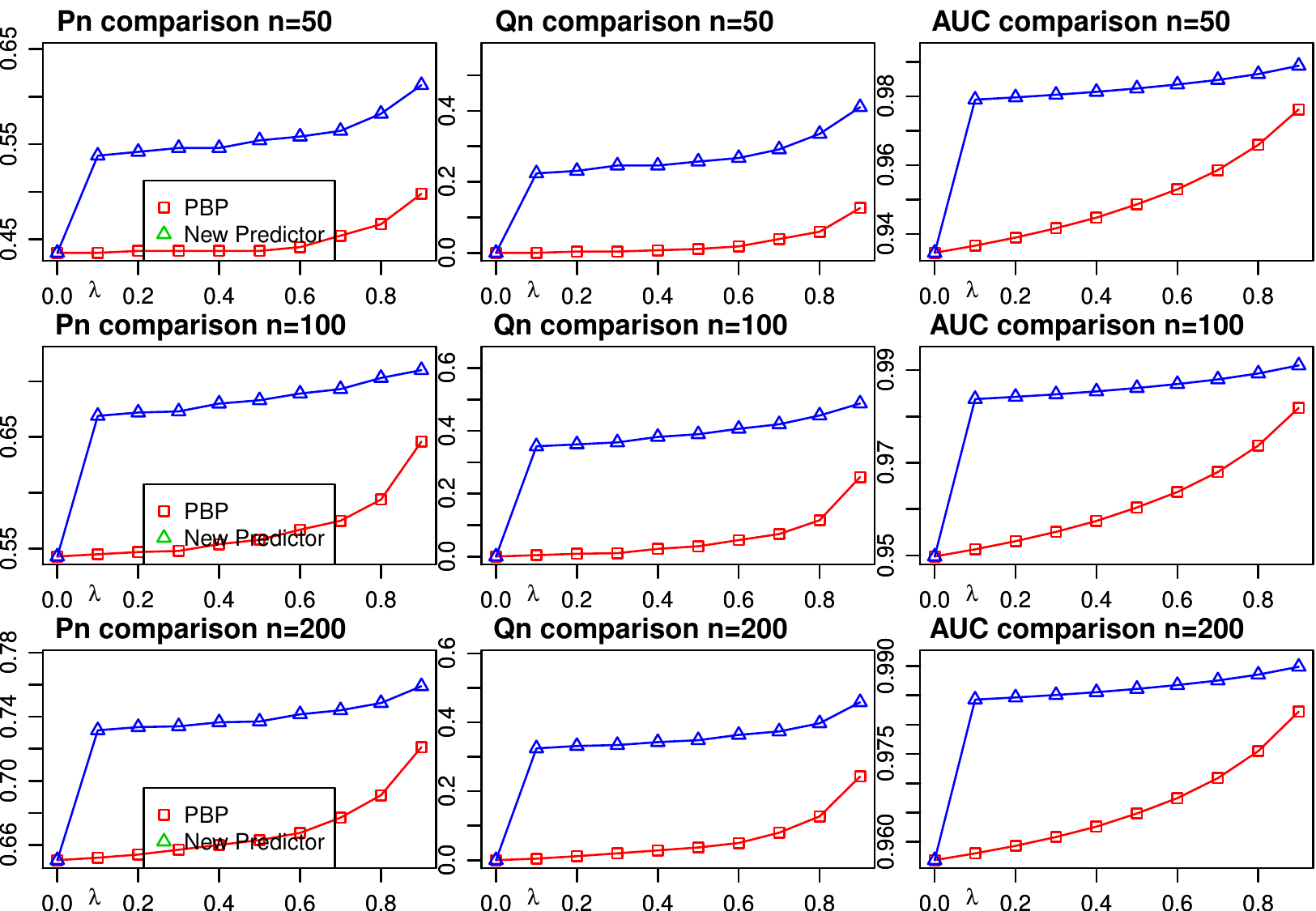} \end{center}

\begin{quote}
Figure 3: The performance comparison between proposed method and base
method for MovieLens dataset. For the above comparison we have
considered \(T_F =30\) and \(T_P =30\) days.
\end{quote}

{[}Figure 3,Figure 4,Figure 5{]} show the comparitive performence of
proposed method over base method. It is easy to find that our proposed
method out performs the base method in all the situation. For all the
datasets \(P_n\) is better for all values of \(n\).\(Q_n\) also out
performs as compare to the base predictor. If the item has low time
span, the prediction made by total popularity is not good while the
prediction made by recent popularity method for the same is good.Results
also show using recent popularity works good for all the datasets;
MovieLens, Netflix and Facebook, and for all the situations. When the
future window length is short.Prediction by total popularity works good
in case of item has already gained long term popularity.

\begin{center}\includegraphics{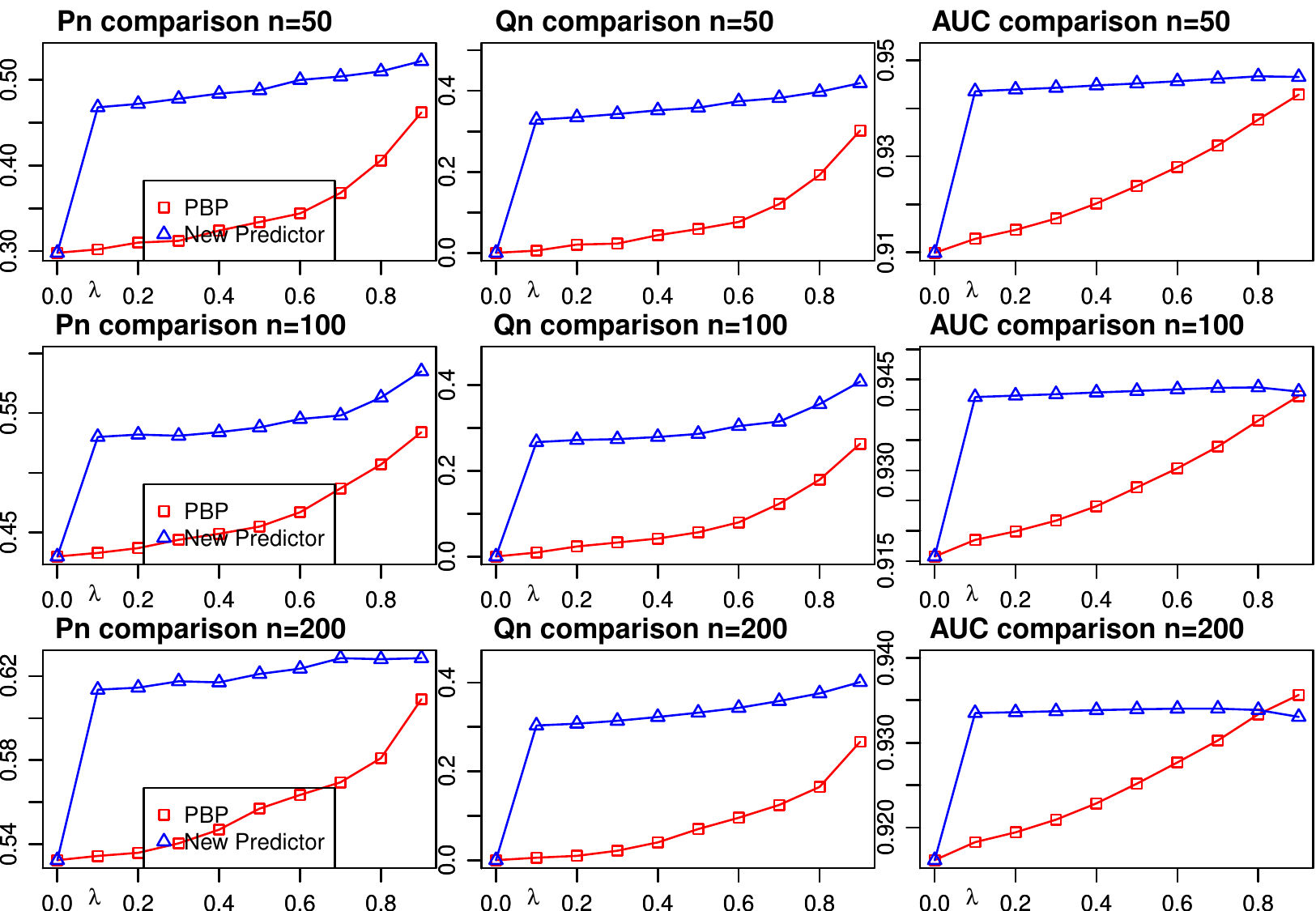} \end{center}

\begin{quote}
Figure 4: The performance comparison between proposed method and base
method for Netflix dataset. For the above comparison we have considered
\(T_F =30\) and \(T_P =30\) days.
\end{quote}

Charecteristric of all the three datasets are not same, such as Facebook
content may not be alive after few weeks while on Movielens and NetFlix
content may never die. Further more the rate at which node attract new
links also differ such as the most popular node will recieve more
attention than less popular ones. The content on Facebook may not be
always intersting to friends so that they can share with their friends
such as friends may not like content on politics. While content on
Movielens and Netflix are appealing to viewers. That is the reason our
predictor's accuracy for Facebook dataset as not good as Movielens and
Netflix. In {[}Figure 6{]} we find precision get better with \(\lambda\)
for all the datasets suggest that people like the item that their peers
are watching or liking in recent time.

\begin{center}\includegraphics{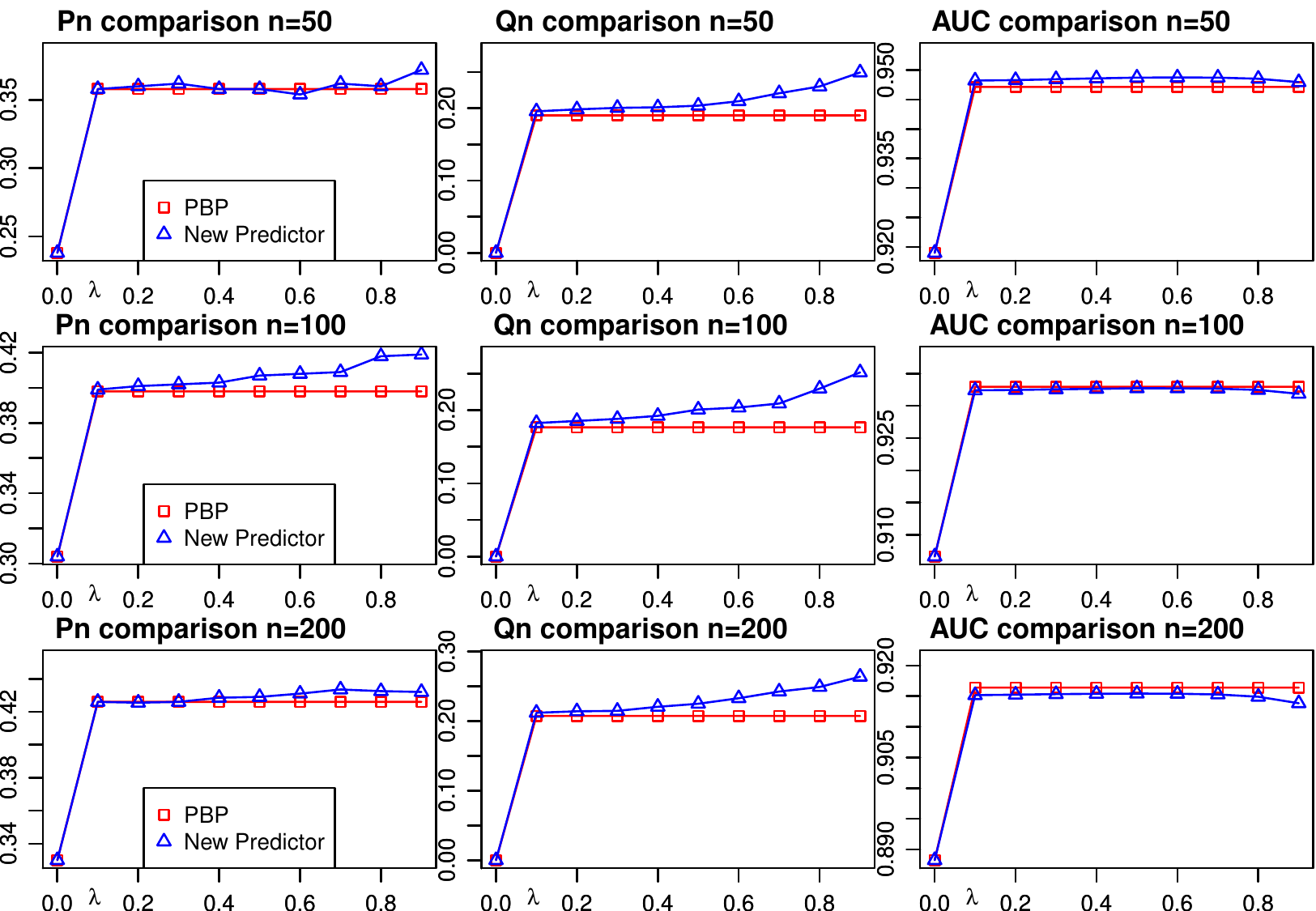} \end{center}

\begin{quote}
Figure 5: The performance comparison between proposed method and base
method for Facebook. For the above comparison we have considered
\(T_F =30\) and \(T_P =30\) days.
\end{quote}

Our predictor also outperforms the base method (based on recent
popularity) also shows that content popularity are affected by it's age
also since we have considered aging in our model. As aging phenomena is
present in every network therefore new entries get chance to become
popular. We can think in two ways, either due to aging of old popular
items new entries get attention or due to quality or fitness of item. If
node achieve popularity due to its fitness suggest it is showing
competitive behaviour.Scientist have found both the phenomena in real
networks. In our case we can argue new entries become popular not only
because old entries aging effect but also items fitness because there
are so many entries to watch or consumed. If people are watching or
liking any item it is because of it's innate quality not because they
dont have enough entries to watch or consume. We know there are plenty
of new entries available for all the cases namely Movielens,Netflix and
Facebook. Therefore we can say that in the presence of quality item they
attract link from the popular item to become popular by showing
competitve behaviour.

\subsection{Predictor's perfomance for varying future time lenght
\(T_F\)}\label{predictors-perfomance-for-varying-future-time-lenght-tux5ff}

In {[}Figure 7{]} we have shown the performance of our predictor against
the base predictor for different values of future time window.For
Proposed predictor \(\lambda = 0.9\) and \(\gamma = 0.1\) for PBP
\(\lambda = 0.9\) and past time window length \(T_P = 60\) days as the
author has used in his paper. It is easy to discover that smaller
\(T_F\) lenght helps in predicting for short time while long \(T_F\)
helps for predicting long term trend.

\includegraphics{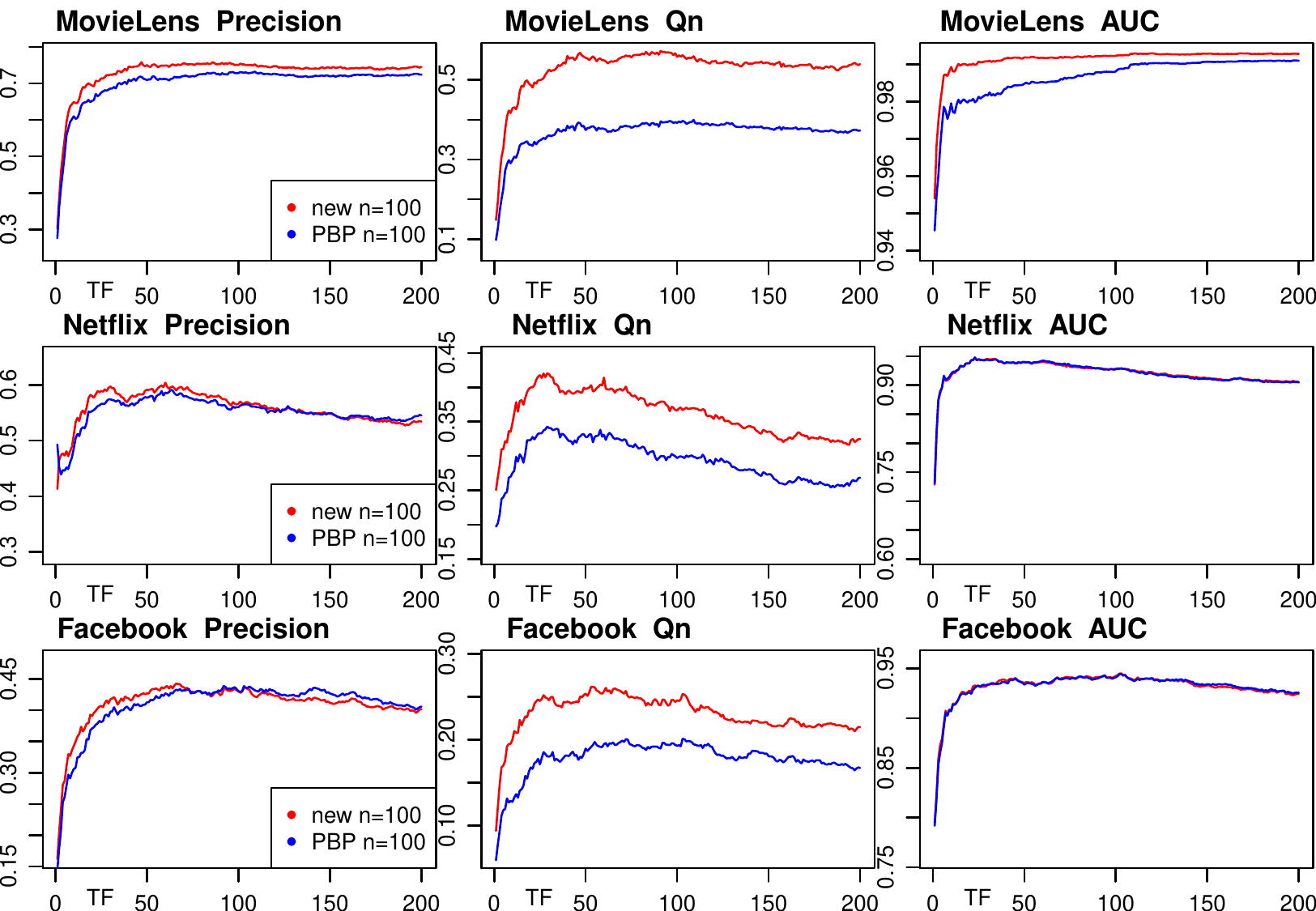}

\begin{quote}
Figure 7: The above figure shows the performance of the predictor for
diferent values on future time window \(T_F\). Red line shows the
performance of our proposed predictor while blue shows the base PBP
predictor.
\end{quote}

We have chosen the \(T_P\) same as \(T_F\) in case of PBP the author has
considered. As we can see from the {[}Figure 7{]} our proposed predictor
has better performance in predicting the long term trend prediction.We
have considered the future time window for \(200\) days to evaluate the
performance of our predictor. We have found that in case of Movielens
people tend to copy the behaviours of their peers as well as they also
like to explore new entries.In case of Netflix we have found precision
shows improvement for around \(100\) days but after that PBP performed
better. Although in enquiry of novel items our predictor is better. This
phenomena also suggest that people explore new itmes on Netlix and also
rely on their peer's recent activity. Facebook activity data shows
similar nature. Our predictor shows significant precision for upto 200
days. It also predicts the behaviour of consumers for exploring or
trying ``new thing'' nature. Our works describe presence ``potantial
items'' those are generally subdued in the presence of other already
popular items.

\section{Conclusion}\label{conclusion}

In this manuscript we came up with one model to make prediction of
object on online social media specially considering it's temporal
behavior.We created a model by considering object's recent popularity as
well as it's aging or decay of popularity.Epirical results show that our
proposed method out performs the base method i.e popularity based
predictor given by (ZENG et al. 2013).We have found that people tend to
copy recent behaviours of their peer consumers not the whole popularity
of items. We have also found that in presence of quality items recent
popular items loses it's popularity or we can say that on these kinds of
network \textbf{competitive behavior} (described by (Bianconi and
Barab{Ã}{\textexclamdown}si 2001) for social network) also found.We have
considered only temporal effects of the node's attracting new link.We
have found it one of the important feature for making prediction.In
future work one can also consider other effects like human dynamics,item
category,node centrality etc.

\section*{References}\label{references}
\addcontentsline{toc}{section}{References}

Asur, Sitaram, Bernardo Huberman, and others. 2010. ``Predicting the
Future with Social Media.'' In \emph{Web Intelligence and Intelligent
Agent Technology (WI-IAT), 2010 IEEE/WIC/ACM International Conference
on}, edited by IEEE, 1:492--99. IEEE.
doi:\href{http://dx.doi.org/10.1109/WI-IAT.2010.63}{10.1109/WI-IAT.2010.63}.

Bianconi, G., and A.-L. Barab{Ã}{\textexclamdown}si. 2001. ``Competition
and Multiscaling in Evolving Networks.'' \emph{EPL (Europhysics
Letters)} 54 (4): 436.
\url{http://stacks.iop.org/0295-5075/54/i=4/a=436}.

Cheng, Justin, Lada A. Adamic, P. Alex Dow, Jon M. Kleinberg, and Jure
Leskovec. 2014. ``Can Cascades Be Predicted?'' In \emph{23rd
International World Wide Web Conference, Www '14, Seoul, Republic of
Korea, April 7-11, 2014}, edited by Chin-Wan Chung, Andrei Z. Broder,
Kyuseok Shim, and Torsten Suel, 925--36. ACM.
doi:\href{http://dx.doi.org/10.1145/2566486.2567997}{10.1145/2566486.2567997}.

Eisler, Zolt{\a'a}n, Imre Bartos, and J{\a'a}nos Kert{\a'e}sz. 2008.
``Fluctuation Scaling in Complex Systems: Taylors Law and Beyond1.''
\emph{Advances in Physics} 57 (1). Informa UK Limited: 89--142.
doi:\href{http://dx.doi.org/10.1080/00018730801893043}{10.1080/00018730801893043}.

et.al, A. Barabasi. 1999. ``Emergence of Scaling in Random Networks.''
\emph{Science} 286 (5439). American Association for the Advancement of
Science (AAAS): 509--12.
doi:\href{http://dx.doi.org/10.1126/science.286.5439.509}{10.1126/science.286.5439.509}.

Gleeson, James P, Davide Cellai, Jukka-Pekka Onnela, Mason A Porter, and
Felix Reed-Tsochas. 2014. ``A Simple Generative Model of Collective
Online Behavior.'' \emph{Proceedings of the National Academy of
Sciences} 111 (29). National Acad Sciences: 10411--15.
doi:\href{http://dx.doi.org/10.1073/pnas.1313895111}{10.1073/pnas.1313895111}.

Herlocker, Jonathan L., Joseph A. Konstan, Loren G. Terveen, and John T.
Riedl. 2004. ``Evaluating Collaborative Filtering Recommender Systems.''
\emph{ACM Transactions on Information Systems} 22 (1). Association for
Computing Machinery (ACM): 5--53.
doi:\href{http://dx.doi.org/10.1145/963770.963772}{10.1145/963770.963772}.

Leskovec, Jure, Lada A Adamic, and Bernardo A Huberman. 2007. ``The
Dynamics of Viral Marketing.'' \emph{ACM Transactions on the Web (TWEB)}
1 (1). ACM: 5.
doi:\href{http://dx.doi.org/10.1145/1232722.1232727}{10.1145/1232722.1232727}.

Leskovec, Jure, Lars Backstrom, and Jon M. Kleinberg. 2009.
``Meme-Tracking and the Dynamics of the News Cycle.'' In
\emph{Proceedings of the 15th ACM SIGKDD International Conference on
Knowledge Discovery and Data Mining, Paris, France, June 28 - July 1,
2009}, edited by John F. Elder IV, Fran{ç}oise Fogelman-Souli{\a'e},
Peter A. Flach, and Mohammed Javeed Zaki, 497--506. ACM.
doi:\href{http://dx.doi.org/10.1145/1557019.1557077}{10.1145/1557019.1557077}.

{Matus Medo Manuel S. Mariani}, An Zeng, and Yi-Cheng Zhang. 2015.
``Identification and Modeling of Discoverers in Online Social Systems.''
\url{http://arxiv.org/pdf/1509.01477.pdf}.
\url{http://arxiv.org/pdf/1509.01477.pdf}.

Onnela, J.-P., and F. Reed-Tsochas. 2010. ``Spontaneous Emergence of
Social Influence in Online Systems.'' \emph{Proceedings of the National
Academy of Sciences} 107 (43). Proceedings of the National Academy of
Sciences: 18375--80.
doi:\href{http://dx.doi.org/10.1073/pnas.0914572107}{10.1073/pnas.0914572107}.

Parolo, Pietro Della Briotta, Raj Kumar Pan, Rumi Ghosh, Bernardo A.
Huberman, Kimmo Kaski, and Santo Fortunato. 2015. ``Attention Decay in
Science.'' \emph{J. Informetrics} 9 (4): 734--45.
doi:\href{http://dx.doi.org/10.1016/j.joi.2015.07.006}{10.1016/j.joi.2015.07.006}.

Szabo, Gabor, and Bernardo A. Huberman. 2015. ``Predicting the
Popularity of Online Content.'' \emph{Communications of the ACM}.
\url{http://doi.acm.org/10.1145/1787234.1787254}.
doi:\href{http://dx.doi.org/10.1145/1787234.1787254}{10.1145/1787234.1787254}.

Tatar, Alexandru, Marcelo Dias {de Amorim}, Serge Fdida, and Panayotis
Antoniadis. 2014. ``A Survey on Predicting the Popularity of Web
Content.'' \emph{J Internet Serv Appl} 5 (1). Springer Science Business
Media.
doi:\href{http://dx.doi.org/10.1186/s13174-014-0008-y}{10.1186/s13174-014-0008-y}.

Yang, Jaewon, and Jure Leskovec. 2011. ``Patterns of Temporal Variation
in Online Media.'' In \emph{Proceedings of the Fourth ACM International
Conference on Web Search and Data Mining}, edited by ACM, 177--86. WSDM
'11. New York, NY, USA: ACM.
doi:\href{http://dx.doi.org/10.1145/1935826.1935863}{10.1145/1935826.1935863}.

ZENG, AN, STANISLAO GUALDI, MAT{\a'U}{Š} MEDO, and YI-CHENG ZHANG. 2013.
``Trend Prediction in Temporal Bipartite Networks: The Case of
Movielens, Netglix, and Digg.'' \emph{Advances in Complex Systems} 16
(04n05). World Scientific Pub Co Pte Lt: 1350024.
doi:\href{http://dx.doi.org/10.1142/s0219525913500240}{10.1142/s0219525913500240}.

Zhu, Han, Xinran Wang, and Jian-Yang Zhu. 2003. ``Effect of Aging on
Network Structure.'' \emph{Physical Review E} 68 (5). American Physical
Society (APS).
doi:\href{http://dx.doi.org/10.1103/physreve.68.056121}{10.1103/physreve.68.056121}.

\end{document}